\documentclass[%
reprint,
 amsmath,amssymb,
 aps,
prab,
]{revtex4-2}

\usepackage{graphicx}
\usepackage{dcolumn}
\usepackage{bm}
\usepackage{algorithm,algcompatible}
\usepackage{textcomp}
\usepackage{siunitx}

\begin{document}

\preprint{APS/123-QED}

\title{Efficient 6-dimensional phase space reconstructions from experimental measurements using generative machine learning}

\author{Ryan Roussel}
\affiliation{SLAC National Accelerator Laboratory, Menlo Park, CA 94025, USA}
\author{Juan Pablo Gonzalez-Aguilera}
\affiliation{Department of Physics and Enrico Fermi Institute, University of Chicago, Chicago, IL 60637, USA}
\author{Eric Wisniewski}
\affiliation{Argonne National Laboratory, Lemont, IL 60439, USA}
\author{Alexander Ody}
\affiliation{Argonne National Laboratory, Lemont, IL 60439, USA}
\author{Wanming Liu}
\affiliation{Argonne National Laboratory, Lemont, IL 60439, USA}
\author{Auralee Edelen}
\affiliation{SLAC National Accelerator Laboratory, Menlo Park, CA 94025, USA}
\author{John Power}
\affiliation{Argonne National Laboratory, Lemont, IL 60439, USA}
\author{Young-Kee Kim}
\affiliation{Department of Physics and Enrico Fermi Institute, University of Chicago, Chicago, IL 60637, USA}

\date{\today}

\begin{abstract}
Next-generation accelerator concepts, which hinge on the precise shaping of beam distributions, demand equally precise diagnostic methods capable of reconstructing beam distributions within 6-dimensional position-momentum spaces. However, the characterization of intricate features within 6-dimensional beam distributions using current diagnostic techniques necessitates a substantial number of measurements, using many hours of valuable beam time. Novel phase space reconstruction techniques are needed to reduce the number of measurements required to reconstruct detailed, high-dimensional beam features in order to resolve complex beam phenomena, and as feedback in precision beam shaping applications. In this study, we present a novel approach to reconstructing detailed 6-dimensional phase space distributions from experimental measurements using generative machine learning and differentiable beam dynamics simulations. We demonstrate that this approach can be used to resolve 6-dimensional phase space distributions from scratch, using basic beam manipulations and as few as 20 2-dimensional measurements of the beam profile. We also demonstrate an application of the reconstruction method in an experimental setting at the Argonne Wakefield Accelerator, where it is able to reconstruct the beam distribution and accurately predict previously unseen measurements 75x faster than previous methods.


\end{abstract}

\maketitle


\section{Introduction}
Current and future particle accelerators rely on the precise control of beam distributions in 6-dimensional position-momentum phase space, especially as accelerators push the boundaries of intensity and brightness for collider and light source applications \cite{nagaitsev_accelerator_2021}.
For example, magnetized electron beams for hadron beam cooling \cite{derbenev_magnetization_1977}, two bunch operations in free electron lasers \cite{marinelli_high-intensity_2015}, and drive/witness bunch pairs for high efficiency wakefield acceleration \cite{lindstrom_energy-spread_2021}, are enabled by the precise control of 6-dimensional phase spaces beyond macroscopic (RMS) beam properties.

Achieving this level of control requires measurement techniques that provide detailed information about the beam distribution in all 6 phase space coordinates, including cross-correlations between the different phase spaces.
Measurements of this type are needed to provide feedback to accelerator operators or autonomous control algorithms to tune accelerator parameters such that the measured beam distribution matches what is needed by applications.
Additionally, detailed characterization of 6-dimensional phase space distributions is necessary to resolve complex beam dynamics phenomena that couples particle motion along multiple axes, such as coherent synchrotron radiation \cite{braun_emittance_2000} or plasma wakefield accelerators \cite{lindstrom_emittance_2022}.


A wide variety of beam manipulation and diagnostic techniques have been developed to measure and predict detailed characteristics of phase space distributions.
These techniques can involve rotating the beam in phase space (tomography) \cite{yakimenko_electron_2003,sander_beam_1980,hermann_electron_2021, mckee_phase_1995}, using masks or meshes to isolate and observe the dynamics of portions of the transverse beam \cite{marx_single-shot_2018}, using specialized, non-destructive beam diagnostics such as laser wires \cite{wong_4d_2022}, or using machine learning models \cite{wolski_transverse_2022,scheinker_adaptive_2021,scheinker_adaptive_2023} to predict the beam distribution from experimental measurements.

Reconstructing detailed 5- or 6-dimensional phase space distributions from experimental data has proven to be substantially more difficult than reconstructing lower dimensional spaces (4 or less phase space coordinates).
Experimental 5-dimensional phase space characterization has been done in limited instances, once at the ARES beamline with a polarizable X-band transverse deflecting cavity \cite{marchetti_experimental_2021,al_5d_2023} using the Simultaneous Algebraic Reconstruction Technique (SART) algorithm \cite{andersen_simultaneous_1984}, and once at the Spallation Neutron Source (SNS) using a set of movable masking slits \cite{hoover_analysis_2023}.
Full 6-dimensional reconstruction of a single beam distribution has also only been done once, by combining multiple, scanning masking slits with a dipole spectrometer and a bunch shape monitor at the SNS beamline \cite{cathey_first_2018}.
However, these measurements required a significant amount beam time resources to carry out (960 measurements over $\sim28$ hrs for the 5-dimensional ARES case, $\sim5$ million measurements over $\sim32$ hrs for the 6-dimensional SNS measurement), making these measurement procedures impractical for regular use as feedback for online accelerator tuning or for understanding complex beam dynamics phenomena.
There is a need for reconstruction methods that significantly reduce the number of measurements required to reconstruct the 6-dimensional beam distribution in order to be used during regular accelerator operations.

In previous work, we introduced and demonstrated a novel method for reconstructing detailed beam distributions, which we refer to here as generative phase space reconstruction (GPSR), using generative machine learning models and differentiable beam dynamics simulations \cite{roussel_phase_2023}.
The generative machine learning model creates nearly arbitrary beam distributions in 6-dimensional phase space by transforming random samples drawn from a fixed probability distribution into macroparticles with realistic position-momentum coordinates.
A differentiable beam dynamics simulation, which allows for analytical calculation of derivatives during each computational step \cite{gonzalez-aguilera_towards_2023}, is then used to train parameters of the generative model (which in turn modifies the generated beam distribution) by minimizing the difference between experimental measurements and simulated predictions.
Combining generative models with differentiable beam dynamics simulations allows the generative model to be trained solely from experimental measurements in a short amount of time, removing the need for large initial training sets usually required by machine learning workflows.
This method was used to reconstruct detailed, correlated, 4-dimensional phase space distributions using a single quadrupole scan (10 measurements of the beam distribution) by taking advantage of the information contained in conventional measurements of the transverse beam profile using a YAG screen \footnote{noting that a quadrupole scan does not provide information about the longitudinal phase space distribution}.
More recently, this method was also used to reconstruct beams with large angular momentum components (magnetization) and large emittance ratios \cite{kim_four-dimensional_2024}.

In this work, we extend the use of GPSR to reconstructing 6-dimensional phase space distributions from experimental measurements.
We demonstrate in simulation that a diagnostic beamline consisting of quadrupoles, a transverse deflecting cavity, and a dipole spectrometer can be used to resolve detailed characteristics of 6-dimensional phase space distributions using as few as 20 measurements.
This reduces the time necessary to reconstruct the beam distribution from many hours to several minutes.
We show that the reconstruction technique accurately reconstructs a variety of different beam distributions including correlated Gaussian distributions, nonlinear distributions, and beam distributions similar to those produced by the emittance exchange (EEX) beamline at the Argonne Wakefield Accelerator (AWA) \cite{ha_precision_2017}.
We then apply the GPSR algorithm to reconstructing a beam distribution from experimental measurements at AWA and demonstrate that it makes accurate predictions of previously unseen measurements.
Finally, we discuss current limitations and advantages of the reconstruction technique.

\begin{figure*}
    \includegraphics[width=\linewidth]{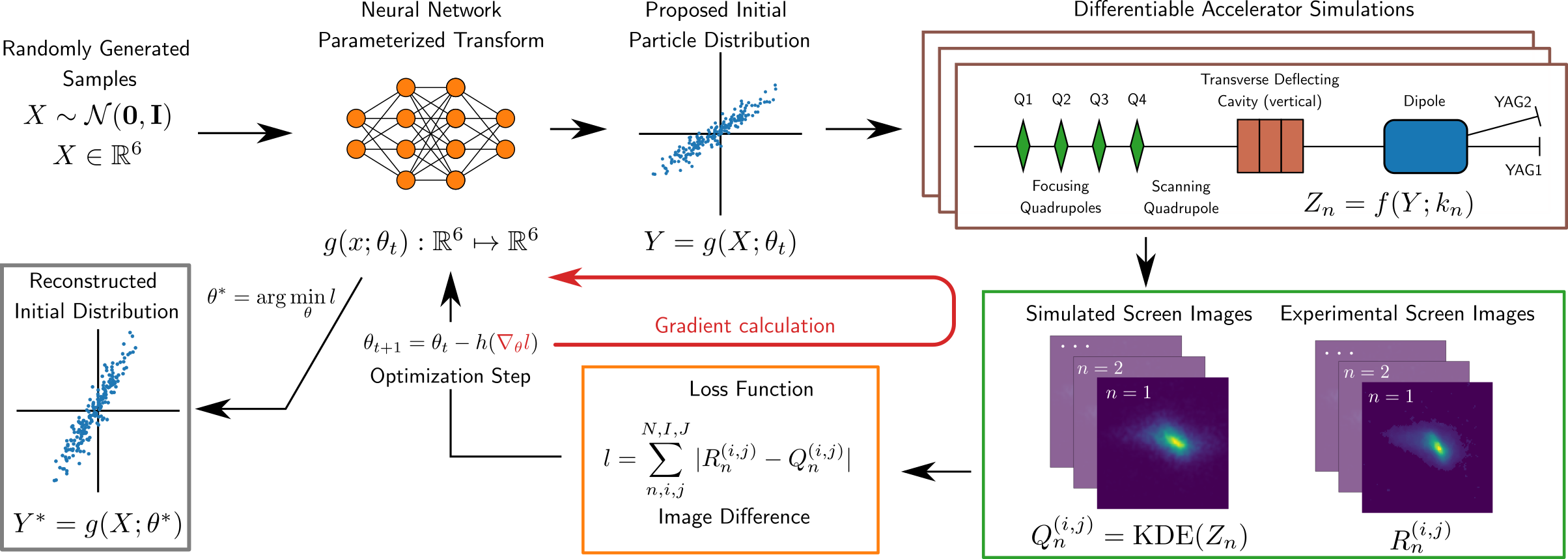}
    \caption{Description of the GPSR approach for reconstructing phase space beam distributions from experimental data. A 6-dimensional beam distribution is parameterized via a generative machine learning model, where randomly generated samples drawn from a Multivariate Normal distribution are transformed using a neural network into macroparticle coordinates in position-momentum space. The beam distribution is then transported through a backwards differentiable simulation of the diagnostic beamline to make measurement predictions at the diagnostic screens. The total per-pixel difference in intensity between simulated predictions $Q^{(i,j)}_n$ and experimental measurements $R^{(i,j)}_n$ is calculated as a loss function. The neural network parameters $\theta$ are then optimized to minimize the loss function using gradients calculated by the differentiable beam dynamics simulation. The distribution generated once the loss function has been minimized (simulation predictions match experimental measurements) is the reconstructed beam distribution.  
    }
    \label{fig:cartoon}
\end{figure*}

\section{Methods}
The method we use to reconstruct 6-dimensional phase space distributions addresses two issues encountered by conventional reconstruction techniques.

First, conventional reconstruction techniques often use 1 or 0-dimensional (scalar) projections of the 6-dimensional phase space to infer features of the distribution.
Conventional algebraic reconstruction techniques such as Maximum Entropy Tomography (MENT) \cite{hock_study_2013} and SART \cite{andersen_simultaneous_1984} typically use 1-dimensional projections of screen images to produce a higher dimensional reconstruction.
Measuring the intensity of the beam distribution through a set of slits, as done in \cite{cathey_first_2018}, further reduces the measurement down to a scalar quantity.
As a result, these methods lose a significant amount of information about the beam distribution, requiring more individual measurements of the distribution to resolve 6-dimensional features.
The GPSR technique enables us to fully utilize the detailed information contained in 2-dimensional images of the beam distribution, without the need to project to lower dimensions.

Second, conventional techniques for representing the distribution of particles in a beam scale poorly when describing  6-dimensional distributions.
Both of the methods listed above use a high-dimensional histogramming approach to describe the beam distribution, i.e., they solve for the beam density inside a number of high-dimensional voxels along an N-dimensional mesh.
While this formalism is used to effectively describe distributions in 1-2 dimensions, it becomes impractical to describe beam distributions in 6-dimensional space this way, as the number of bins grows exponentially with the number of dimensions.
For example, resolving a beam distribution with 100 bins per dimension results in $10^{12}$ voxels needed to describe the full 6-dimensional distribution.
Alternative methods for representing beam distributions, such as distribution moments or principal component analysis techniques, reduce reconstruction detail.

To address these challenges, GPSR introduces two new concepts shown in Fig.~\ref{fig:cartoon} to describe and solve for detailed 6-dimensional phase space distributions.

First, we use a generative machine learning model to represent a distribution of macro particles in 6-dimensional phase space.
This is inspired by the development of neural radiance fields \cite{mildenhall_nerf_2020}, which uses neural networks to represent mass and color density functions in 3-dimensional space.
GPSR uses a generative model, in this case a small neural network, to transform randomly generated samples from a 6-dimensional multivariate Normal distribution into real macroparticle phase space coordinates.
Neural networks of sufficient complexity are universal function approximators \cite{hornik_multilayer_1989}, enabling the generative model to produce particle distributions with nearly arbitrary structure in 6-dimensional phase space.
For the work here, we found that a fully connected neural network with 2 layers, 20 neurons each, connected by Tanh activation functions, was sufficient to represent nearly arbitrary beams with high enough detail.
With this method, the distribution of macroparticles in phase space is entirely controlled by the parameters of the neural network, resulting in a parameterization of the distribution using approximately 1000 free parameters, as opposed to the millions of parameters needed for mesh-based representations of 6-dimensional distributions.

The reconstruction process determines the parameters of the generative beam model by solving an optimization problem, where the goal is to minimize the mean squared error between simulated predictions and experimental measurements, in this case the per-pixel intensity of the transverse YAG screens.
Due to the number of free parameters contained inside the generative beam distribution model, solving this optimization problem in a reasonable amount of time is beyond the capabilities of black box optimization algorithms because of the so-called ``curse of dimensionality" \cite{bellman_dynamic_1966}.

Given a large number of free parameters, it is necessary to use gradient-descent-based algorithms to solve the GPSR optimization problem.
Unfortunately, this is also prohibitively difficult when using conventional beam dynamics simulations to predict experimental measurements, since calculating the gradients numerically in these cases requires finite-difference methods that also scale poorly with the number of free parameters used in optimization.

To address this challenge, GPSR leverages ``backwards-mode" \cite{lecun_efficient_2012} automatic differentiation to substantially reduce the cost of evaluating gradients of simulation outputs with respect to input parameters.
Automatic differentiation is the technique of tracking derivative information alongside each computation step during physics simulations.
This in turn, allows analytical evaluation of simulation output derivatives by using the chain rule in reverse to determine derivatives, as process commonly referred to as ``backpropagation" \cite{lecun_efficient_2012} or ``adjoint-differentiation" \cite{dorigo_toward_2023}. 
Computational costs of calculating the derivatives in this manner is roughly equivalent to the cost of evaluating the simulation itself, and more importantly, is \textit{independent} of the number of input parameters the derivative is taken with respect to.
This makes calculating derivatives substantially cheaper to compute than finite-difference methods when optimizing with respect to a large number of input parameters.
It should be noted that this process is similar to, but distinct from, previous uses of automatic differentiation in accelerator physics, often referred to as ``forward-mode" differentiation \cite{dorigo_toward_2023}, ``differential algebra" \cite{martin_differential_1999}, or, when computing higher derivatives ,``truncated power series algebra" \cite{deniau_generalised_2015}.
These techniques are well suited for calculating particle transport dynamics up to arbitrary order, however they do not scale well to calculating gradients with respect to thousands of input parameters.

Facilitating the use of backpropagation in the context of GPSR requires beam dynamics and diagnostic simulations that support the tracking of derivatives during evaluations.
To this end, we have developed the simulation package \textit{Bmad-X} \cite{gonzalez-aguilera_towards_2023} which re-implements beam transport through a number of simple accelerator elements using the machine learning library \textit{PyTorch} \cite{paszke_pytorch_2019} which implements backpropagation.
Additionally, we simulate the measurement of transverse beam profile intensity on a screen diagnostic by using kernel density estimation (KDE) \cite{weglarczyk_kernel_2018} as opposed to normal histogramming to preserve differentiability.
By using differentiable beam dynamics simulations, we are able to cheaply compute derivatives for use in gradient descent optimization of beam distribution parameters to reconstruct the beam distribution.
More recently, similar beam dynamics simulation packages have been developed, namely the Cheetah Python package \cite{kaiser_cheetah_2024}, although they were not used in this work.

\begin{figure}
    \includegraphics[width=\linewidth]{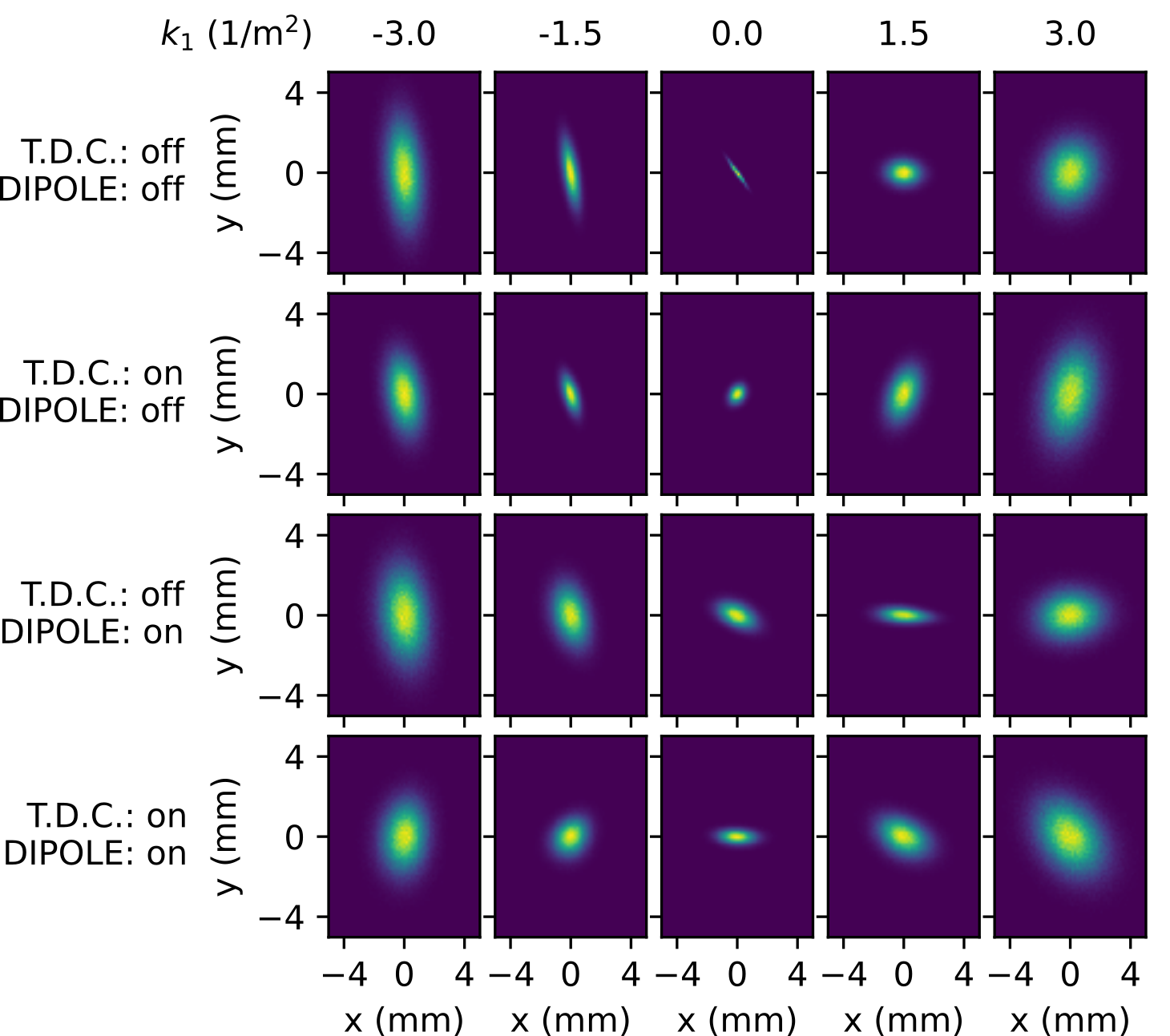}
    \caption{Simulated screen images of the Gaussian beam distribution during the 6-dimensional reconstruction scan. Brighter colors denote higher beam intensity (arbitrary scale for each image).}
    \label{fig:gaussian_measurements}
\end{figure}

\section{6-dimensional diagnostic beamline description}
In previous work \cite{roussel_phase_2023}, we demonstrated that this reconstruction technique is able to produce accurate predictions of the 4-dimensional beam distribution using images gathered from a single quadrupole scan.
To extend this work towards resolving 6-dimensional phase spaces, we add a transverse deflecting cavity (TDC) and a dipole spectrometer to the diagnostic beamline shown in Fig.~\ref{fig:cartoon}.
In this case, quadrupoles Q1-Q3 are used to focus the beam onto YAG1 and quadrupole Q4 is scanned to measure the transverse phase space distribution.
Focusing the beam on the diagnostic screen improves the measurement resolution and increases the range of Q4 strengths that can be scanned over since the transverse beam profile needs to be kept within a region of interest on both of the diagnostic screens.
Pairing a TDC that kicks the beam vertically to resolve the current profile of the beam with a horizontally bending dipole magnet, which measures beam energy spread, is a common approach taken to measure the longitudinal phase space distribution \cite{gao_single-shot_2018}.
The diagnostic setup used in this work is motivated by the notion that combining transverse information from the quadrupole scan with the longitudinal phase space manipulations should provide enough information to resolve the full 6-dimensional phase space distribution.

We demonstrate the 6-dimensional reconstruction technique using a simulation of the diagnostic beamline at AWA.
Transverse diagnostic screens YAG1 and YAG2 are placed along each beam path when the dipole is off and on respectively.
These simulated screens have a region of interest that is 5 x 5 mm (or 200 x 200 pixels) in size, with a resolution of 25 $\mu$m/px.
The L-band transverse deflecting cavity used in simulation (1.3~GHz, $L=0.48$~m) operated with a peak field in a range from 0-3~MV depending on the beam distribution, consistent with the operational range of the AWA deflecting cavity \cite{conde_commissionning_2012}.
The rectangular dipole spectrometer ($L=0.3018$~m) has a bend angle of 20 degrees.

\begin{figure*}
    \includegraphics[width=\linewidth]{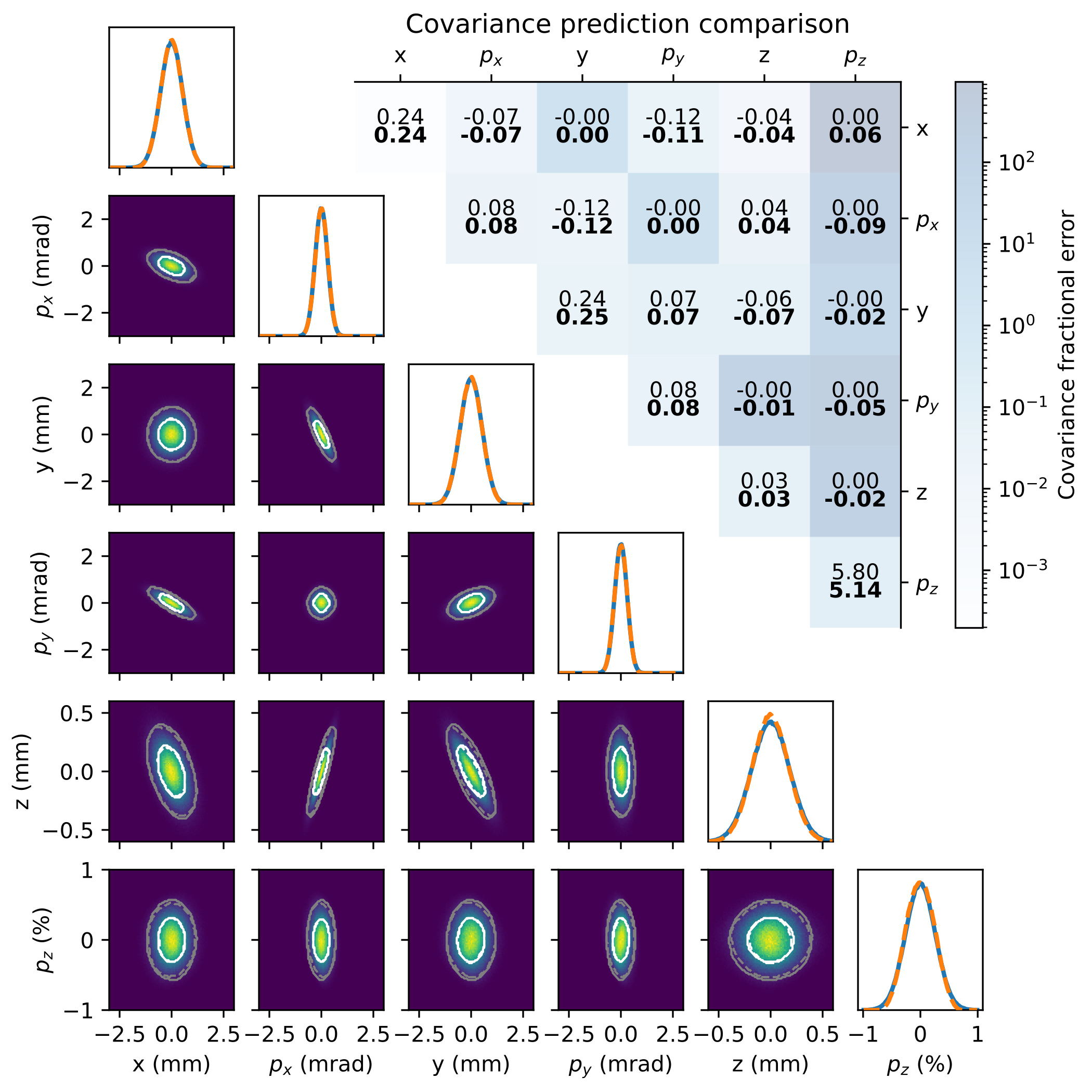}
    \caption{Reconstruction results from Gaussian synthetic beam distribution. Lower left: Comparison between 2-dimensional projections of the ground truth synthetic beam distribution with the reconstructed beam distribution. Solid lines denote ground truth projections and contours while dashed lines denote reconstruction predictions. White and grey contours denote 50th and 90th percentile intensity levels. Color map intensity denotes reconstructed prediction. Upper right: Comparison between ground truth and reconstructed second order moments of 90th percentile beam particles, where regular text denotes ground truth values and bold text denotes reconstructed predictions.}
    \label{fig:gaussian_phase_space_prediction}
\end{figure*}

\section{Synthetic reconstruction Examples}
\label{sec:synthetic_reconstructions}
In this section, we evaluate the effectiveness of the GPSR technique on several different phase space distributions.
For each case we describe below, the following procedure was followed.
The first three quadrupoles (Q1, Q2, Q3) are tuned using Bayesian optimization implemented in Xopt \cite{roussel_xopt_2023} to minimize the transverse beam size at YAG1 with the scanning quadrupole (Q4) turned off.
We then scan the focusing strength of Q4 four times, once for each combination of TDC and dipole states.
Images from the quadrupole scans are then used to reconstruct the beam distribution.
In the synthetic case studies that follow, the quadrupole strength was scanned over five steps, resulting in a total data set of 20 images.

\subsection{Case 1: Gaussian beam reconstruction}
\label{sec:gaussian_reconstruction}
The first case we explore is one where the beam distribution is a Multivariate Normal distribution that contains cross correlations between a number of the 6-dimensional phase space coordinates.
Simulated measurements of the beam on the two diagnostic screens for each of the quadrupole scans is shown in Fig.~\ref{fig:gaussian_measurements}.
These images were then used to reconstruct the beam distribution using the GPSR approach with 100k macro particles.
To maximize reconstruction accuracy, the reconstruction was trained for 3000 iterations of gradient descent (Adam \cite{kingma_adam_2017}, learning rate 0.01) which took roughly 10-15 minutes on an NVIDIA A100 GPU, although its possible that fewer iterations could be used.

The reconstruction algorithm results in a generative model which generates a distribution of macro-particles in 6-dimensional phase space that should approximate the true beam distribution.
The reconstructed beam distribution, with a comparison to the synthetic ground truth distribution, is shown in Fig.~\ref{fig:gaussian_phase_space_prediction}. 
We see from the 50th and 90th percentile contours (calculated by measuring the beam intensities that account for 50/90 percent of the total beam distribution) that the reconstructed beam distribution closely matches the true synthetic distribution.
Furthermore, we can also compare measurements of the second order moments of the beam distribution, which also show close agreement between the reconstruction and the ground truth.
In most cases the fractional error of the reconstructed second order moments is below the 10\% level.
In cases where the cross-covariances of the ground truth distribution are much smaller than the principal axis covariances the error of the reconstruction is higher, due to the dominance of the beam size along the principal axis over other beam features.

\subsection{Case 2: Nonlinear beam reconstruction}
\label{sec:nonlinear_reconstruction}
We now analyze a case where the beam has a variety of non-linearities and correlations throughout the phase space distribution, as shown in Fig.~\ref{fig:non_linear_reconstruction}
 (simulated measurements of this beam distribution can be found in Appendix~\ref{sec:training_data}).
The reconstruction successfully replicates the ground truth distribution generated by performing 4 quadrupole scans with different states of the longitudinal phase space diagnostic elements, including nonlinear features contained in cross-correlations of the beam distribution.
Looking at the comparison between the predicted and ground truth values of the covariance matrix again shows good agreement within 10\% for most elements.

\begin{figure*}
    \includegraphics[width=\linewidth]{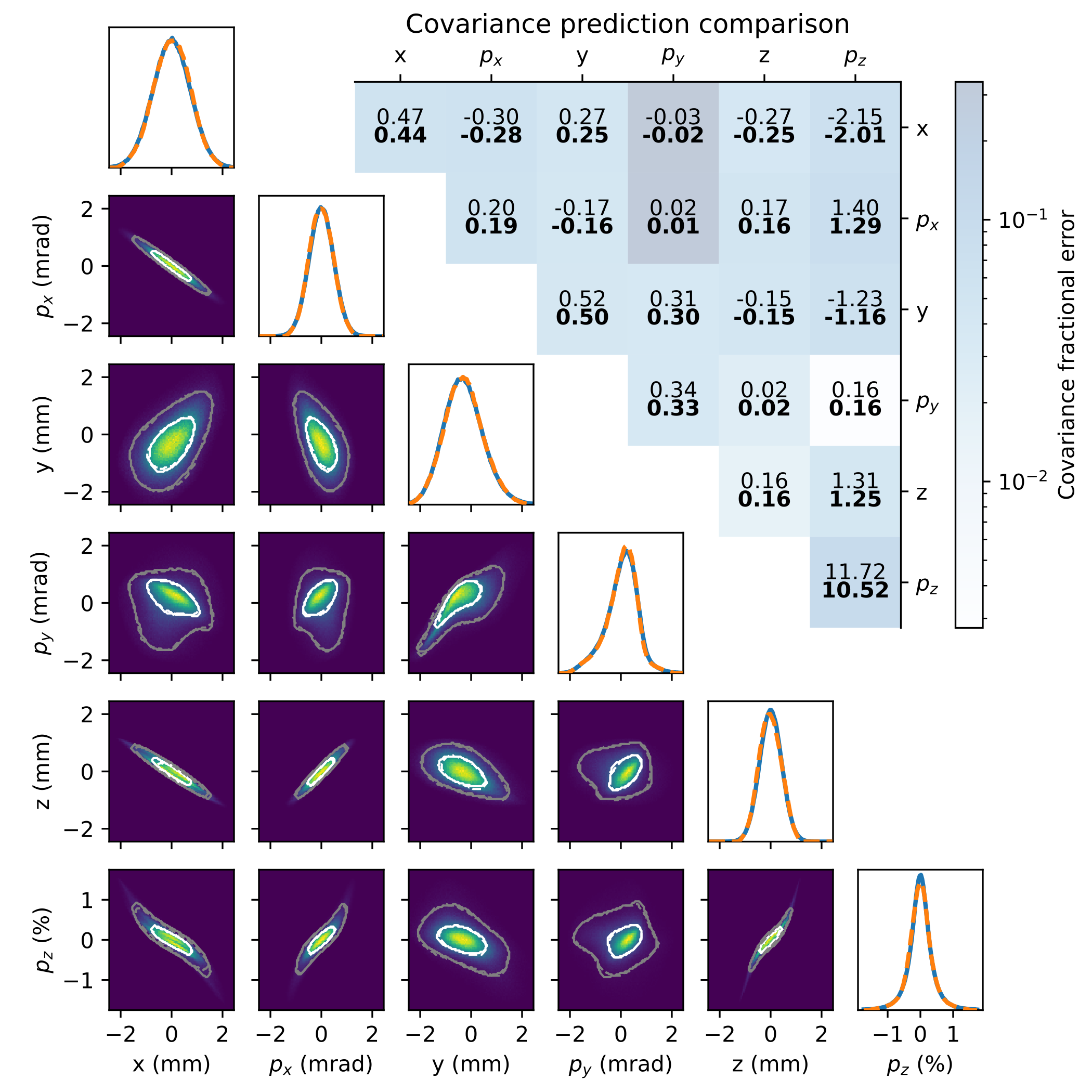}
    \caption{Reconstruction results from nonlinear synthetic beam distribution. Lower left: Comparison between 2-dimensional projections of the ground truth synthetic beam distribution with the reconstructed beam distribution. Solid lines denote ground truth projections and contours while dashed lines denote reconstruction predictions. White and grey contours denote 50th and 90th percentile intensity levels. Color map intensity denotes reconstructed prediction. Upper right: Comparison between ground truth and reconstructed second order moments of 90th percentile beam particles, where regular text denotes ground truth values and bold text denotes reconstructed predictions.}
    \label{fig:non_linear_reconstruction}
\end{figure*}




\subsection{Case 3: EEX beam reconstruction}
\label{sec:eex_reconstruction}
Detailed 6-dimensional phase space reconstructions are particularly important when performing complex beam manipulations that impart significant correlations in the 6-dimensional phase space distribution and contain precisely shaped features.
An example of this is transverse-to-longitudinal emittance exchange (EEX) \cite{cornacchia_transverse_2002,ha_precision_2017}.
This process combines two doglegs and a transverse deflecting cavity to map the horizontal phase space distribution into a longitudinal distribution, allowing longitudinal profile shaping using transverse masking.
For example, the EEX beamline has been used to generate ramped current profiles with a sharp drop off to improve the transformer ratio of dielectric \cite{gao_single-shot_2018} and plasma \cite{roussel_single_2020} wakefield acceleration.
Precisely characterizing beam distributions created by the EEX beamline enables us to control accelerator and beam parameters, such as the location of leafs in a multi-leaf masking element \cite{majernik_beam_2023} or focusing magnets before the EEX beamline \cite{ha_perturbation-minimized_2016}, that lead to improved 6-dimensional beam tailoring for accelerator applications.

As a proof-of-concept demonstration, we generated a synthetic beam distribution which mimics those created by the EEX beamline for high transformer ratio wakefield acceleration applications.
In this case, EEX aims to create a drive beam with a triangular current profile and a uniform witness beam to sample the wakefield.
This is achieved using a laser cut mask to shape the horizontal beam profile before EEX, which is then mapped into a current distribution during the exchange process.
However, this results in a correlation between the vertical beam size and longitudinal position within the bunch, since the vertical distribution is not exchanged like the horizontal distribution inside the EEX beamline.
This can be detrimental to achieving optimal matching into a wakefield structure due to time-of-flight degradation of the longitudinal profile inside strong final-focus magnets before the wakefield device.

\begin{figure*}
    \includegraphics[width=\linewidth]{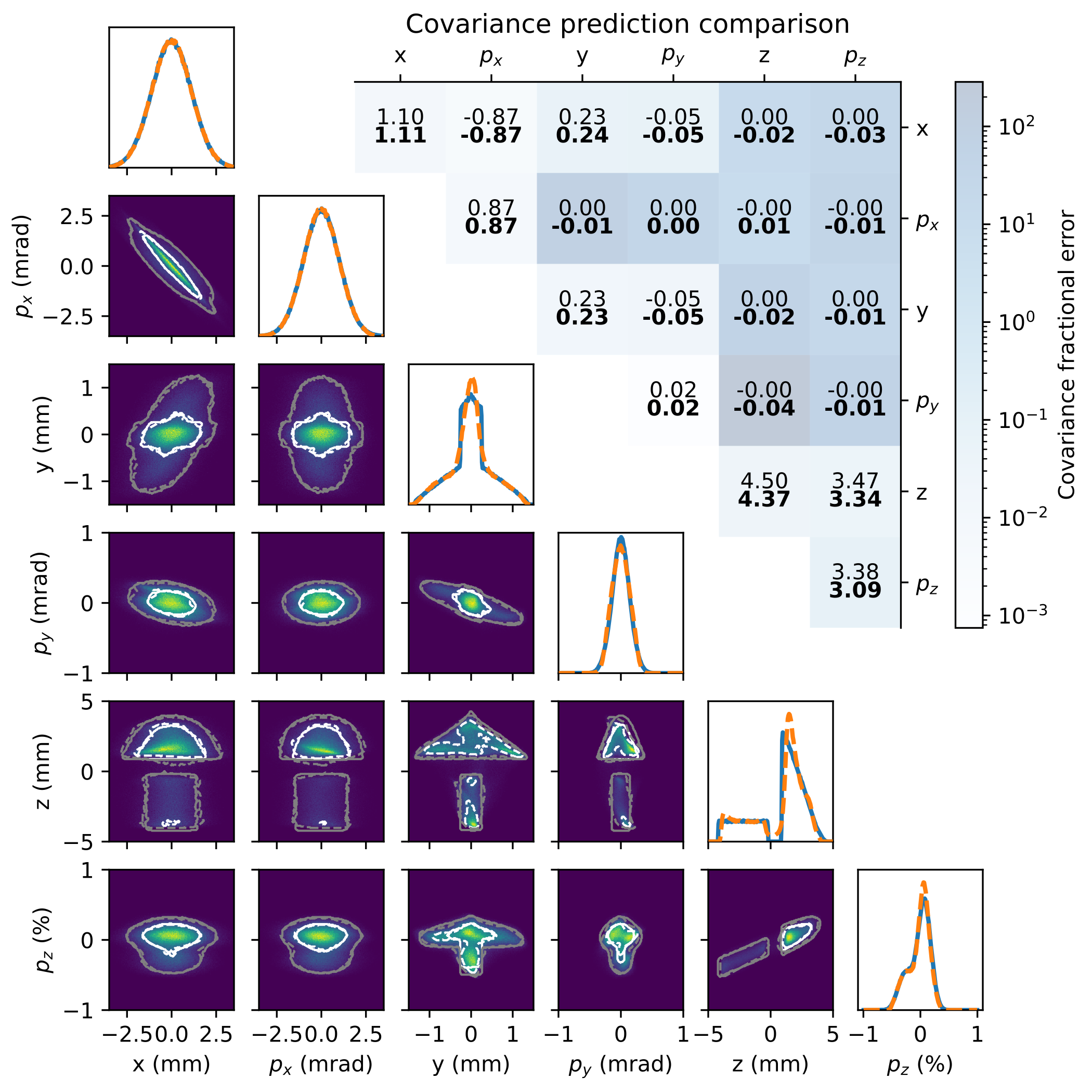}
    \caption{Reconstruction results from the EEX synthetic beam distribution. Lower left: Comparison between 2-dimensional projections of the ground truth synthetic beam distribution with the reconstructed beam distribution. Solid lines denote ground truth projections and contours while dashed lines denote reconstruction predictions. White contours denote 90th percentile intensity levels. Color map intensity denotes reconstructed prediction. Upper right: Comparison between ground truth and reconstructed second order moments of 90th percentile beam particles, where regular text denotes ground truth values and bold text denotes reconstructed predictions.}
    \label{fig:eex_reconstruction}
\end{figure*}

Reconstruction results from simulated diagnostic measurements (seen in Appendix~\ref{sec:training_data}) of this beam distribution are shown in Fig. \ref{fig:eex_reconstruction}.
We see that the reconstruction algorithm can resolve important features of the beam distribution, including the longitudinal profile and correlations between transverse and longitudinal phase spaces.
However, the reconstruction algorithm has difficulty identifying large areas of uniform density within the beam, most notably in the $z-y$ phase space where the triangular head and the lower rectangular regions should both have near uniform density profiles.
This leads to slightly inaccurate predictions of the longitudinal beam profile, which is critical for high transformer ratio applications.
Identifying how to improve the accuracy of the reconstruction in this case, either through algorithm modifications or changes in the diagnostic beamline, is a topic of future study.

\section{Experimental Demonstration}
\label{sec:exp_reconstruction}
In addition to simulation studies, we conducted an experimental demonstration of using the 6-dimensional GPSR algorithm to characterize electron beams generated at AWA.
We configured the AWA drive beamline to produce 1~nC electron bunches at a 2~Hz repetition rate with an energy of 43~MeV and transported them to the diagnostic configuration shown in Fig.~\ref{fig:cartoon} at the end of the AWA beamline.
The beam charge was selected to provide a strong enough signal-to-noise ratio for beam imaging diagnostics while mitigating coherent synchrotron radiation effects in the dipole spectrometer.

Quadrupoles Q1-3 were then used to focus the beam onto YAG1 using Bayesian optimization algorithms while the scanning quadrupole Q4 was turned off.
We then repeated 4 quadrupole scans (-2.9 T/m to 2.9 T/m, 9 steps) with different TDC and dipole settings, as was done in the simulated examples.
A set of 5 beam image shots were taken for each parameter setting with a charge window of 0.1~nC.
The four parameter scans took approximately 8 minutes to perform at AWA, although we estimate that given better charge stability, the scan time could be reduced by a factor of 2.

An additional complexity of the experimental measurement was a slight difference in image resolution between the two YAG screens used to measure the beam profile due to minor differences in camera location and focusing.
These differences were incorporated into the reconstruction by defining two beamline simulations; one for the case where the dipole was off, and one case where the dipole was on, each with different definitions for the screen diagnostic.
The images were cropped to a square size of 300 x 300 pixels, which corresponds to a side length of approximately 13 mm.
Finally, the intensity of the image was clipped and set to zero at a lower bound threshold and a Gaussian smoothing filter was applied to the images to remove salt and pepper noise.

To mitigate shot-to-shot jitter of the beam, we shifted the distribution of each image such that the beam centroid was at the center of the region of interest and averaged the intensity profile over the 5 shots.
As a result, our reconstruction predicts the structure of the distribution, but not its offset with respect to magnetic element centers, mean energy, or timing relative to the zero-crossing of the TDC cavity.
A more robust treatment of the shot-to-shot jitter would allow us to identify these aspects of the beam distribution, as was done in \cite{roussel_phase_2023}, and is a topic of future study.

In order to validate the accuracy of the reconstruction in the absence of a ground truth beam distribution, we compared model predictions to a subset of data that was not included in determining the phase space distribution (refereed to here as the test data set).
If the generative model can accurately predict measurements inside the test set, we can have confidence that the reconstructed distribution is accurate.
We selected every other quadrupole strength in the 4 quadrupole scans to be test data, resulting in a training data set consisting of 20 images (5 quadrupole settings x 4 LPS diagnostic settings) and a test data set consisting of 16 images (4 quadrupole settings x 4 LPS diagnostic settings).
The entire data set can be viewed in Appendix \ref{sec:training_data}.
With this training data set the reconstruction took approximately 17 minutes on an A100 Nvidia GPUs due to the increased image size compared to simulated examples.

\begin{figure*}
    \includegraphics[width=\linewidth]{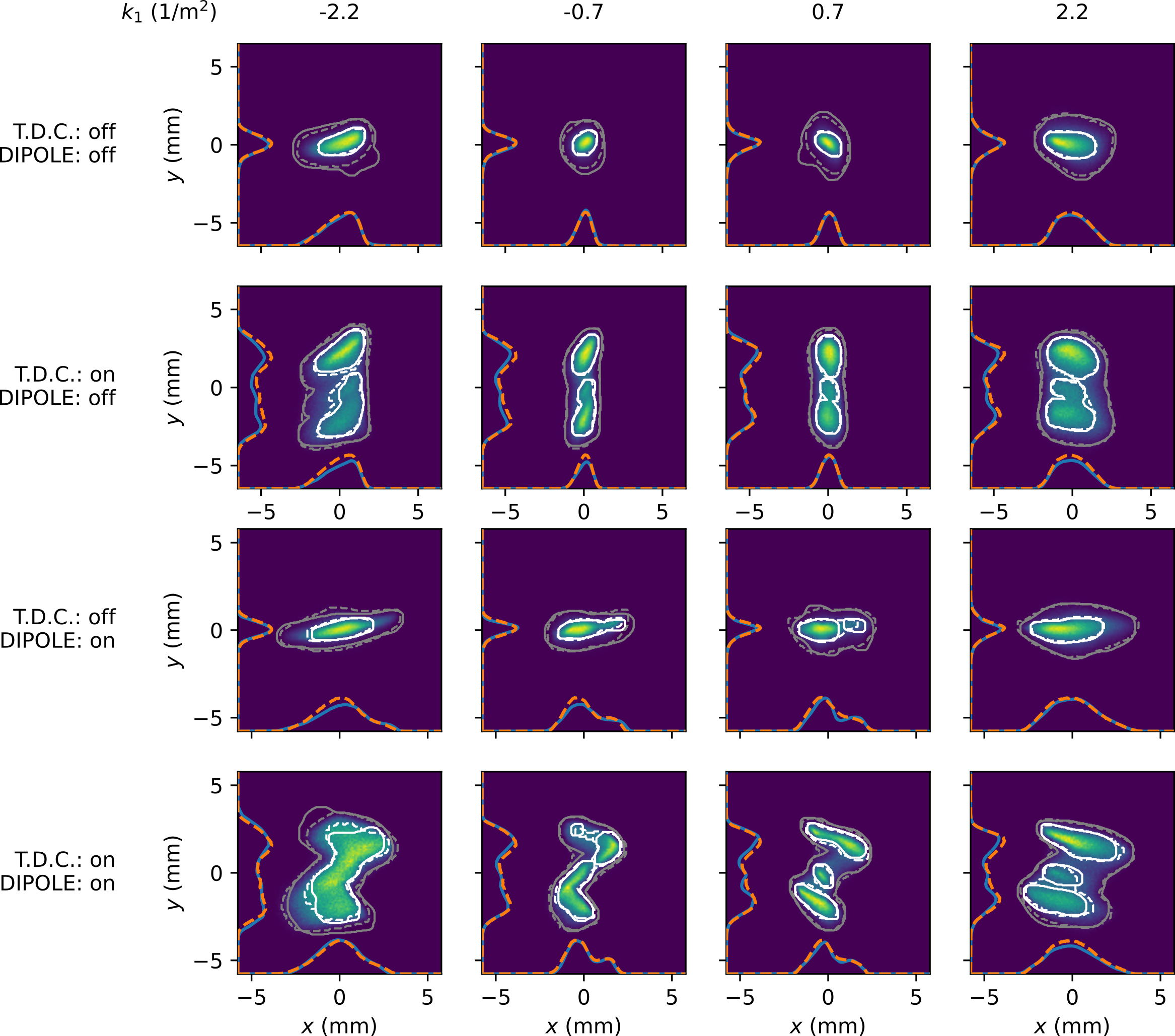}
    \caption{Comparison between averaged, experimentally measured test images with predictions from propagating the reconstructed beam distribution to the screen diagnostics. White and grey contours denote 50th and 90th percentile intensity levels, with solid lines representing measurements and dashed lines denoting predictions. Color map intensity denotes reconstructed prediction on an arbitrary scale. Blue lines denote measured 1-dimensional projections, while orange dashed lines denote projection predictions.}
    \label{fig:exp_prediction_comparison}
\end{figure*}

\begin{figure*}
    \includegraphics[width=\linewidth]{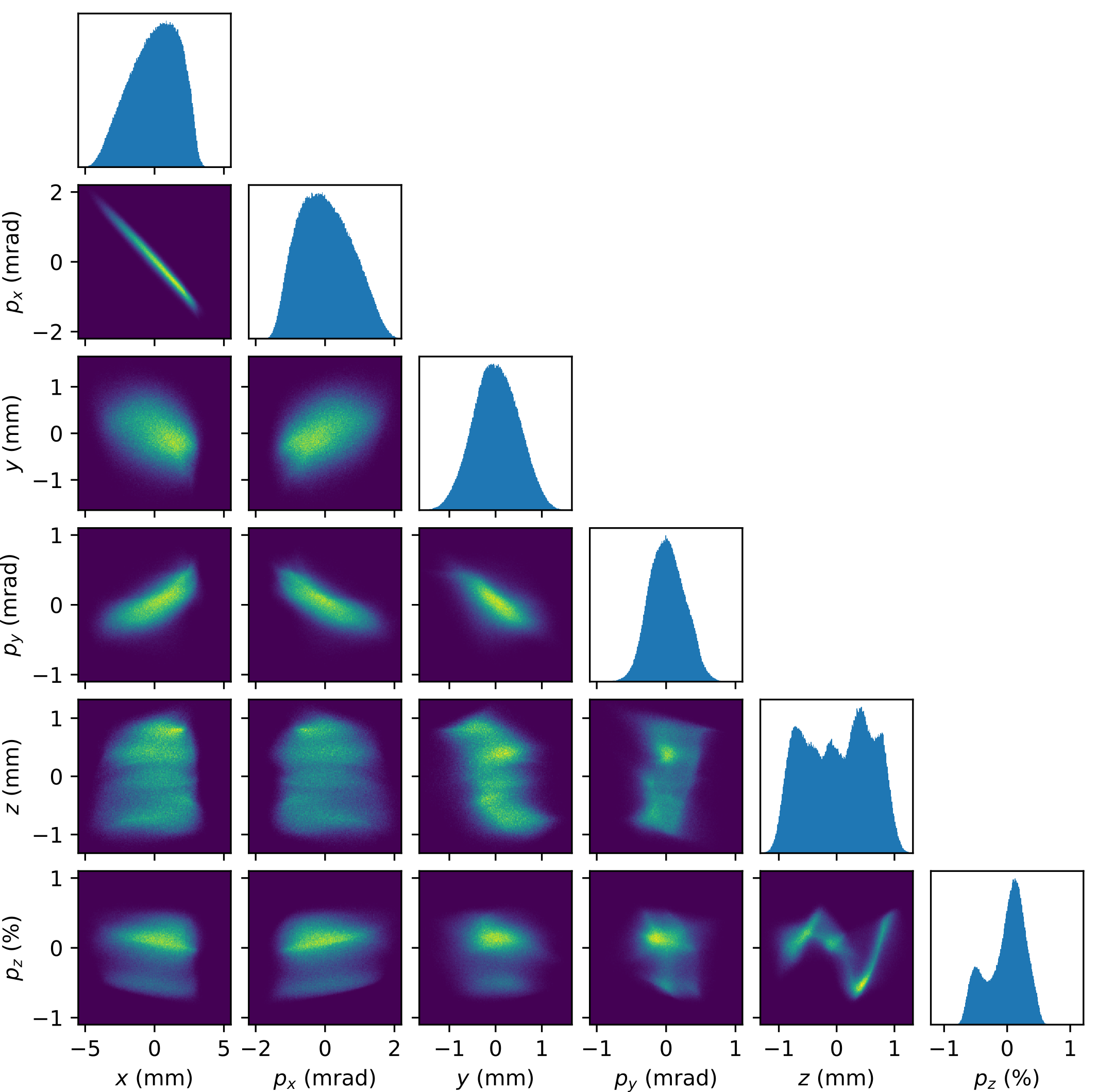}
    \caption{Projections of the reconstructed 6-dimensional beam distributions where the color map intensity denotes beam density (independent scaling for each projection).}
    \label{fig:exp_reconstruction}
\end{figure*}

A comparison between predictions from the reconstruction and the experimentally measured test data is shown in Fig.~\ref{fig:exp_prediction_comparison}.
We see that the reconstruction accurately reproduces the 1 and 2-dimensional beam structure (with minor discrepancies in some cases) seen in the test images, including nonlinear and beamlet features in some cases.
The reconstructed 6-dimensional phase space from experimental measurements is shown in Fig.~\ref{fig:exp_reconstruction}.
The reconstruction predicts normalized transverse emittances of $\varepsilon_{x,n} = 19$ mm-mrad, $\varepsilon_{y,n}= 8.5$ mm-mrad which are consistent with non-optimized AWA beamline parameters, the bunch charge of 1~nC, and the observed asymmetric beam size growth as a function of quadrupole strength when the dipole and deflecting cavity are off.
One prominent feature of the distribution is the beamlet structure observed in the longitudinal current profile.
We believe that this structure results from a set of 5 alpha-BBO crystals used at AWA to produce longer, flat-top laser pulse profiles by stacking 32 laser sub-pulses together \cite{power_temporal_2009}, however it is unclear how this leads to variations in beam energy along the bunch.

\section{Discussion}
While the examples shown here demonstrate the effectiveness of GPSR in these cases, the reliability and robustness of the algorithm can be improved for general applications.
The most critical next step for this work is to enable the algorithm to quantify uncertainties in the reconstructed phase space distribution from limited experimental measurements.
Conventional wisdom and empirical evidence \cite{hock_beam_2011} suggests that evenly-spaced angular rotations of the distribution over 360 degrees is necessary to accurately reconstruct 2-dimensional phase space distributions from 1-dimensional projections.
However, it is not clear how this rule-of-thumb translates to reconstructing beam distributions in 6-dimensional phase space from 2-dimensional images or in the case where nonlinear beam manipulations are used.
Developing an understanding of reconstruction uncertainty would help explain why a given collection of measurements would be sufficient to reconstruct a particular beam distribution.
Additionally, this would aid in further reducing the number of measurements needed to reconstruct a particular beam distribution by tracking the reduction in reconstruction uncertainty as measurements are added to prevent redundant measurements.

The flexibility of the GPSR algorithm can be used to develop novel beam manipulations and diagnostic techniques that were previously impractical to analyze using conventional techniques.
Unlike other analysis methods, the GPSR algorithm does not require simplifications or approximations of beam dynamics or measurement signals needed to enable analytical tractability.
By utilizing high-performance numerical optimization techniques, GPSR can solve extremely complex optimization problems that can incorporate exact measurement information into the reconstruction process.
As a result, GPSR can also easily combine data from multiple, potentially heterogeneous sources of information (as done here, where different diagnostics have different resolutions) about the beam distribution by adding additional terms into the training loss function shown in Fig.~\ref{fig:cartoon}.
For example, screen measurements of the 2-dimensional beam profile can be easily combined with upstream measurements of the beam profile, non-destructive measurements of the beam distribution, such as edge radiation in bends \cite{fiorito_noninvasive_2014}, or ML based predictions of the longitudinal phase space \cite{emma_machine_2018} into a single, self-consistent description of the beam distribution.

Finally, one advantage of GPSR is that the generative model creates these macro-particles by transforming samples from a simple random distribution.
As a result, predicted particle distributions can contain any number of macro-particles regardless of the number of particles used during training the model. 
For example, even though the generative model in the cases demonstrated here was trained with 100k particles, the figures shown here uses predictive distributions containing one million macro-particles.
This method for representing beam distributions is advantageous in a number of ways for use in simulations.
Saving and transporting high-fidelity beam distributions containing a large number of 6-dimensional macro-particle coordinates is a memory intensive process.
Generative models on the other hand use substantially fewer scalar quantities (the weights and offsets of neural network parameters) and a description of the model structure to represent the beam distribution at any level of fidelity, significantly reducing the memory needed to share the beam distribution between simulations.

\section{Conclusion}
In this work, we have demonstrated in several simulated and experimental case studies that detailed characterizations of 6-dimensional beam distributions can be achieved rapidly using the GPSR algorithm.
This analysis method leverages the detail contained in 2-dimensional screen images of the beam distribution and knowledge of beam dynamics in the accelerator to significantly reduce (by up to a factor of 75x) the time needed to produce detailed predictions of the 6-dimensional phase space distribution.
Furthermore, the generative ML representation of the beam distribution is trained from scratch on experimental data, requiring no previous data collection or pre-training needed by other applications of ML in accelerator physics.
As a result, the GPSR algorithm can be used to provide 6-dimensional phase space information during accelerator operations and in a wide variety of contexts.
This technique has major implications for allowing 6-dimensional information to be used to inform accelerator control and understand complex physical phenomena.

\begin{acknowledgments}
Conceptualization, R.R., J.P.G.A., and A.L.E.; Data curation, R.R. and J.P.G.A.,; Formal analysis, R.R., and J.P.G.A.; Funding acquisition, A.L.E., J.P., and Y.K.K.; Investigation, R.R. and J.P.G.A.; Methodology, R.R., J.P.G.A., and A.L.E.; Software, R.R. and J.P.G.A.; Experiment, R.R., J.P.G.A., W.L., E.W., and A.O.; Supervision, A.L.E., R.R., Y.K.K., and J.P.; Validation, R.R., and J.P.G.A.; Visualization, R.R., and J.P.G.A.; Writing---original draft, R.R., and J.P.G.A.; Writing---review and editing, R.R., J.P.G.A., and A.L.E. All authors have read and agreed to the published version of the manuscript.

This work is supported by the U.S. Department of Energy, Office of Science under Contract No. DE-AC02-76SF00515 and the Center for Bright Beams, NSF award PHY-1549132.
This research used resources of the National Energy Research Scientific Computing Center (NERSC), a U.S. Department of Energy Office of Science User Facility located at Lawrence Berkeley National Laboratory, operated under Contract No. DE-AC02-05CH11231 using NERSC award ERCAP0020725.

\end{acknowledgments}

\appendix

\section{Synthetic and Experimental Datasets}
\label{sec:training_data}
\begin{figure*}
    \includegraphics[width=\linewidth]{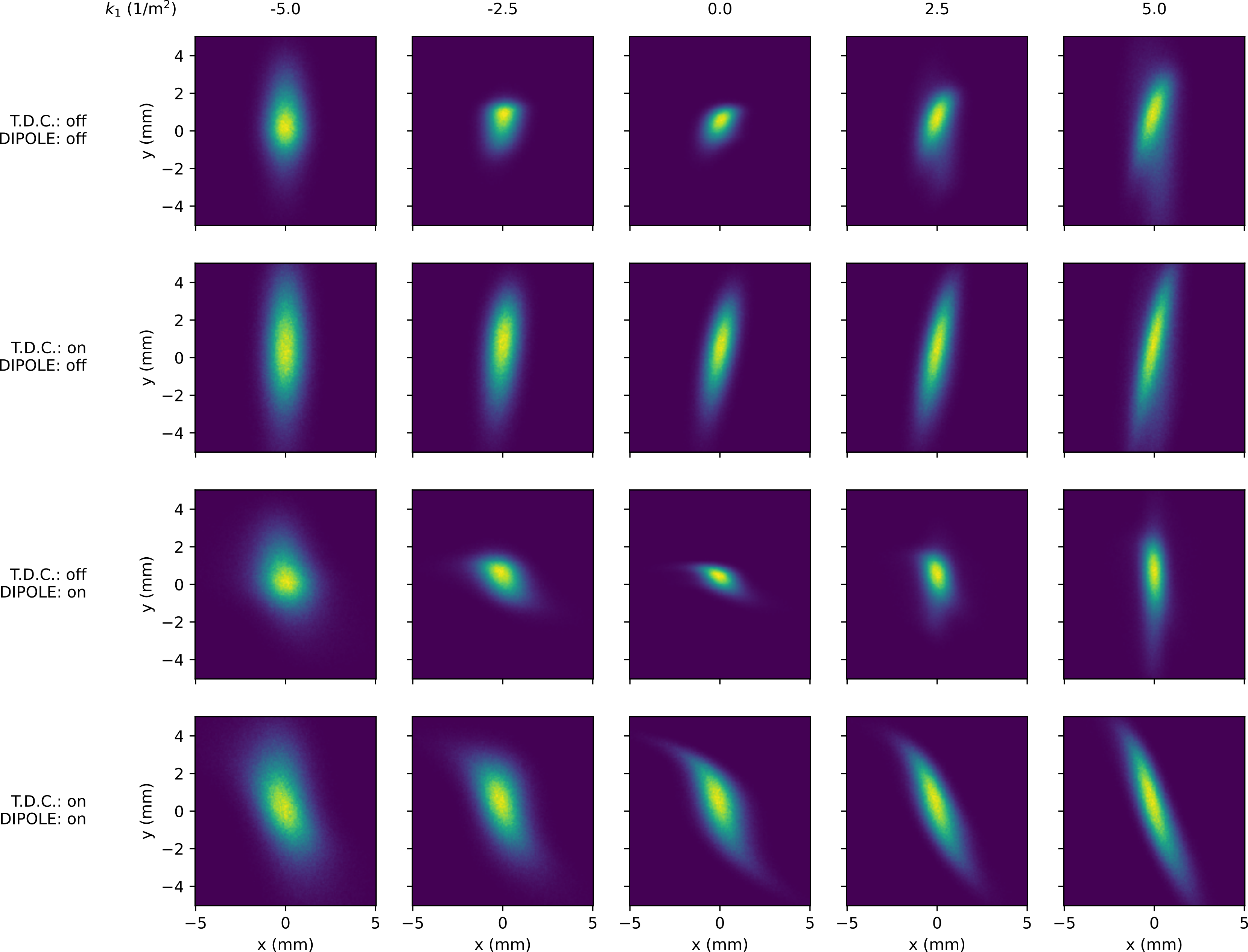}
    \caption{Simulated screen images of the nonlinear beam distribution during the 6-dimensional reconstruction scan. Brighter colors denote higher beam intensity (arbitrary scale for each image).}
    \label{fig:nonlinear_measurements}
\end{figure*}

\begin{figure*}
    \includegraphics[width=\linewidth]{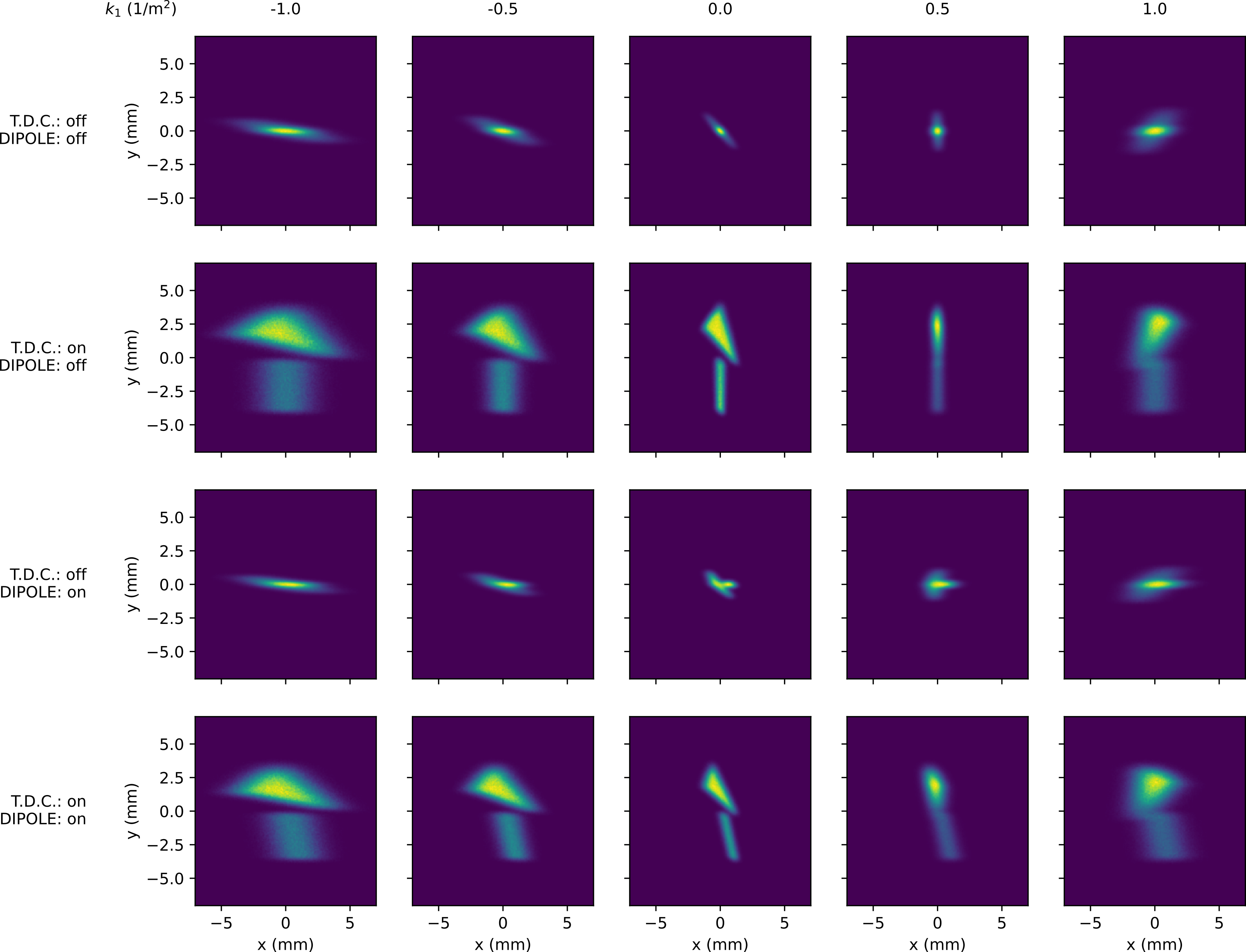}
    \caption{Simulated screen images of the nonlinear beam distribution during the 6-dimensional reconstruction scan. Brighter colors denote higher beam intensity (arbitrary scale for each image).}
    \label{fig:eex_measurements}
\end{figure*}

\begin{figure*}
    \includegraphics[width=\linewidth]{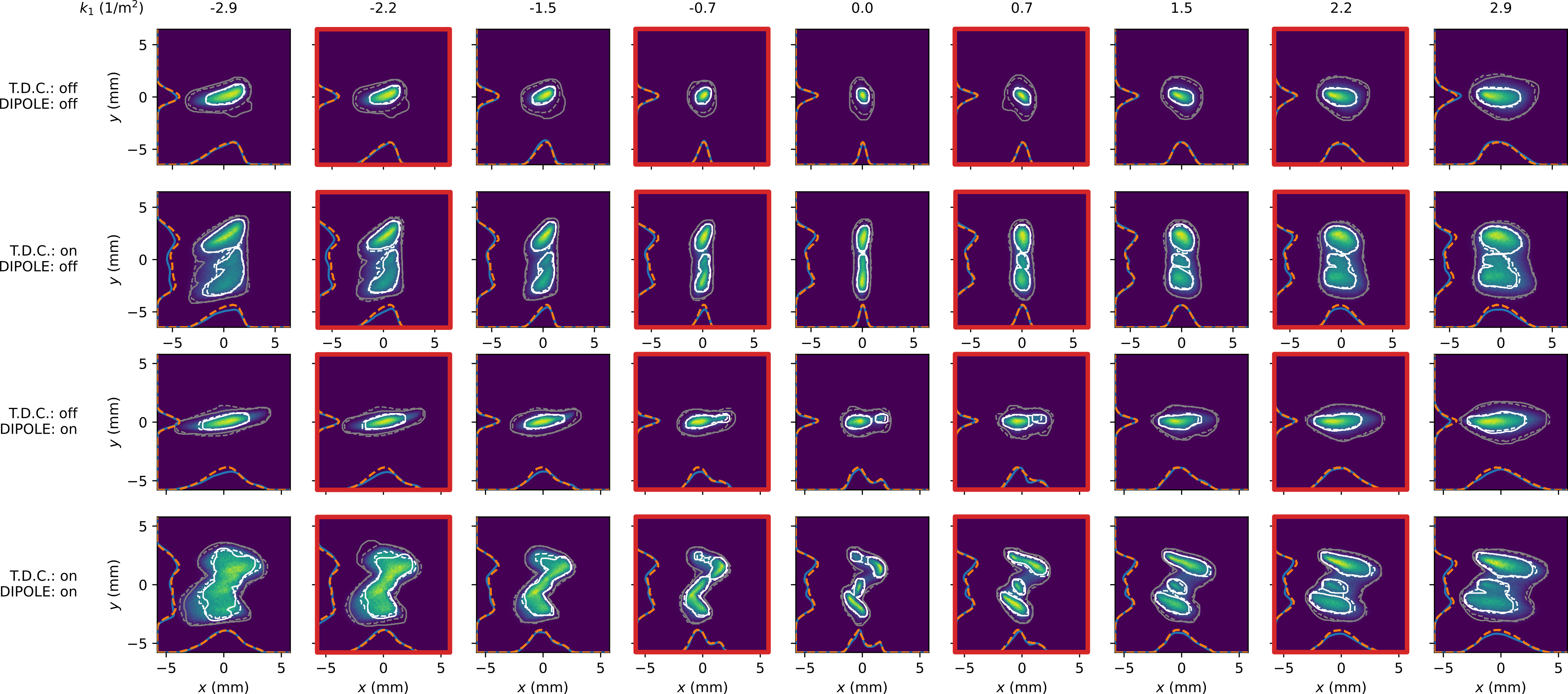}
    \caption{Comparison between all averaged, experimentally measured images with predictions from propagating the reconstructed beam distribution to the screen diagnostics. White and grey contours denote 50th and 90th percentile intensity levels, with solid lines representing measurements and dashed lines denoting predictions. Color map intensity denotes reconstructed prediction on an arbitrary scale. Blue lines denote measured 1-dimensional projections, while orange dashed lines denote projection predictions. Images with red borders denote test images not included in determining the phase space distribution.}
    \label{fig:exp_measurements}
\end{figure*}

Figures~\ref{fig:nonlinear_measurements},~\ref{fig:eex_measurements}, and~\ref{fig:exp_measurements} contain training data sets used to reconstruct the beam distributions described in Section~\ref{sec:synthetic_reconstructions} and Section~\ref{sec:exp_reconstruction} .


%

\end{document}